\documentclass[longbib,aip,rsi,reprint,superscriptaddress,floatfix]{revtex4-1}

\usepackage{graphicx} 
\usepackage{amsmath,amssymb}
\usepackage{xcolor,tabularx}
\usepackage{float}
\usepackage[english]{babel}
\usepackage[autostyle]{csquotes}
\usepackage{xfrac} 
\usepackage{siunitx}
\usepackage{tikz}
\usepackage{wrapfig}
\usepackage{sidecap}

\newcommand{\ee}{\ensuremath\mathrm{e}} 
\newcommand{\dd}{\mathrm{d}} 

\newcommand{\bdm}{\begin{displaymath}}
\newcommand{\edm}{\end{displaymath}}

\renewcommand{\eqref}[1]{Eq.~(\ref{#1})}

\begin{document}

\title{Jamming and Flow in Granular Matter: A Physics Lab Course Experiment} 

\author{Thomas Blochowicz}\affiliation{Institute for Condensed Matter Physics, Technical University of Darmstadt, D-64289 Darmstadt,Germany} 
\author{Emina Ismajli}\affiliation{Institute for Condensed Matter Physics, Technical University of Darmstadt, D-64289 Darmstadt,Germany}
\author{Jan Philipp Gabriel}\email{Jan.Gabriel@dlr.de}\affiliation{German Aerospace Center (DLR), Institute of Frontier Materials on Earth and in Space --  Functional, Granular, and Composite Materials, 51170 Cologne, Germany} 

\begin{abstract}
We describe a dynamic light scattering setup that uses diffusing wave spectroscopy (DWS) to investigate the dynamics in sand grains subjected to periodic vertical shaking by a loudspeaker. Along with the setup that is used in the undergraduate physics lab course at TU Darmstadt, the necessary DWS theory is introduced, including the proper treatment of the oscillatory excitation. Some exemplary results are presented that demonstrate the similarity of jamming in an athermal granular medium with the glass transition in thermally driven molecular systems, a relation that has frequently been pointed out but still is poorly understood. Similar, albeit more sophisticated experiments are currently conducted in microgravity environments such as the international space station ISS and the experiment may serve as an introduction to an exciting field of  current research. 
\end{abstract}

\maketitle
\section{Introduction}

A granular medium is an accumulation of solid particles with a minimum size of approximately \SI{10}{\micro\meter}. Brownian molecular motion is then negligible, and the dynamics of the particles is significantly influenced by gravity in addition to the interaction of the particles with each other. Macroscopic grain collisions are generally inelastic and, in order to keep the particles in motion, a granular system must be continuously supplied with energy, for example, in the form of shaking, pouring, or blowing in a stream of air \cite{Duran2000a}. It is essential that the number of particles considered is large enough so that the exact number no longer plays a role. For this reason, the properties of the system must be recorded as statistical variables, which is why the concepts and experimental methods of statistical mechanics play a key role in the following.
	
The size and homogeneity of the accumulation of particles can vary greatly. We often encounter grains in different variants in everyday life. So many phenomena that occur are quite familiar to us, such as the ``Brazil nut effect'', in which particles of different sizes separate under the influence of vibration (muesli in a jar) \cite{Gajjar2021a,Rosato1987a}, or the formation of ripples on sand dunes \cite{Nishimori1993a,Charru2013a}. On the other hand, there are also seemingly simple questions that still remain difficult to answer \cite{Watson2023a}: Why does sand appear, for example, as a kind of solid (after all, you can walk on it), but in an hourglass as a kind of liquid? Or why does salt only flow out of the shaker when we tip it far enough? 

\begin{figure}[t!]
\centering
\includegraphics[width=0.95\linewidth]{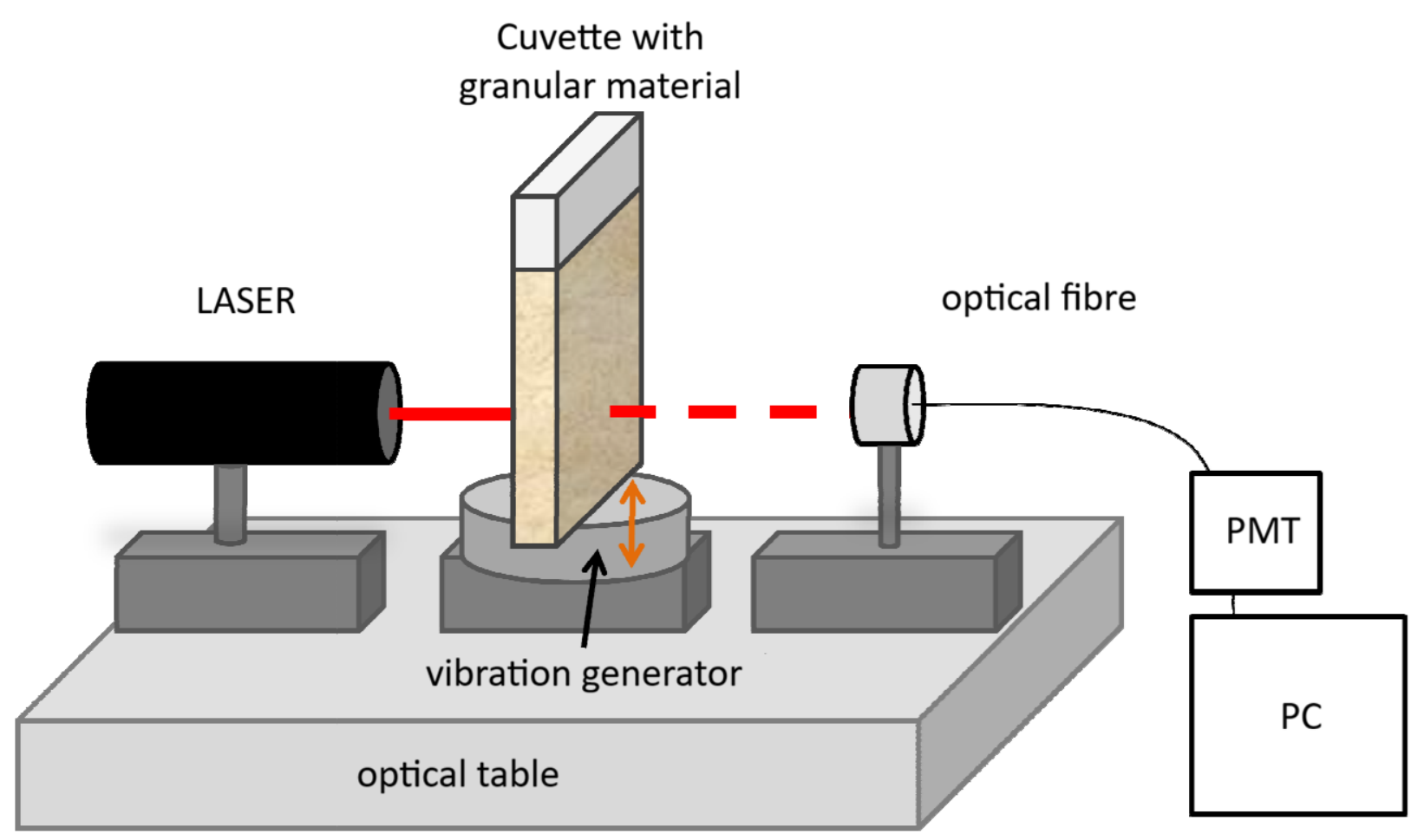}
\caption{The setup used for the DWS experiment \cite{Radetinac2010}.}
\label{fig:aufbau}
\end{figure}

To learn more about granular materials, we will measure the dynamics of granular particles in a setup shown in Fig.~\ref{fig:aufbau} as used in a lab-course designed at TU Darmstadt \cite{M12,Radetinac2010} similar to experiments for loose grains \cite{Pine1988,fraden1990multiple,zivkovic2009study} and more dense systems \cite{mayo2025observing,kunzner2025dynamics}. The particular setup uses laser light to measure the motion of particles shaken by a voice coil, similar to the experiment suggested by Kim et al.\ \cite{Kim2005a}.

\subsection{Granular media and jamming}

The aim is to investigate the jamming transition in a granular system. Jamming refers to the fact that a granular medium can initially be deformed and flow, similar to a conventional liquid, but with only a slightly higher packing density, the granular particles can block each other and prevent further movement. This is, for example, the reason why a dry pile of sand does not simply flow apart or why peanuts can block the hole in a bag, even if the opening is significantly larger than the diameter of a single nut.\cite{Behringer2018a}

In analogy to the dynamic behavior of molecules in different states of aggregation, a distinction is also made in granular systems between different phases: solid -- liquid -- gas, due to the mobility of the particles but with the difference that in grains the mobility is not caused by thermal energy, as in molecular systems, but by the influence of external forces such as vibration, a gas flow or similar \cite{Andreotti2013a}. The term vibration-fluidized bed is used to express the fact that in grains excited by vibrations, the movement of the grain particles is comparable to the Brownian motion of particles in a molecular liquid due to the frequent collisions between the particles.

The ``jammed state'', on the other hand, is characterized by the disorder contained in the system. Firstly, the system is structurally disordered, much like the molecules in a liquid. There is also another type of disorder: jamming in grains is based on the transfer of forces at the contact points between neighboring particles. However, because the particles are never identical and spherical, there is a disorder of contact forces because a particle typically has many more nearest neighbors than it needs contact points to stabilize its position \cite{Behringer2018a}.

\begin{figure}[t!]
    \centering
    \includegraphics[width=0.8\linewidth]{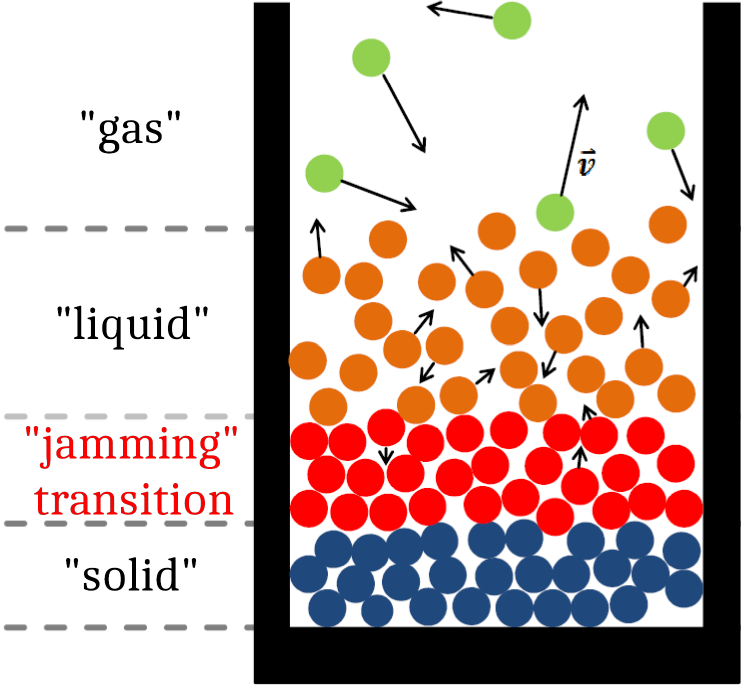}    
    \caption{Different states of motion in a granular system caused by the variation of the packing density or by the effect of external forces such as vibration. The dynamics are often compared with the states of aggregation in molecular systems: bottom: glassy (amorphous) solidified, solid; middle: liquid; top: gaseous. Between ``solid'' and ``liquid'': the jamming transition. \cite{Radetinac2010}}
    \label{fig:gran-dyn}
\end{figure}

Therefore, regardless of the structural arrangement, the actual transmission of mechanical forces is limited to a few contact points, and thus, the direction of force transmission is random and disorganized. In this way, force-transmitting contact points form chains and overall a network of so-called ``force chains'', which run through the system in a completely disordered manner \cite{vHecke2005a}.

In this way, force chains ultimately transfer forces to the base or walls of the vessel containing the granules. Even if the system is considerably stabilized along the direction of local stress, a small force perpendicular to that stress is often sufficient to loosen contact points and thus release the blockage. It is therefore a good idea to shake a packet of peanuts that is stuck to get hold of the coveted food. Such systems, in which a small force perpendicular to the direction of stress can release the blockage, are known as ``fragile matter''.\cite{Cates1998a} However, not all systems that exhibit a jamming transition are ``fragile'' in this way. Especially not if the dynamic blockage is accompanied by considerable elastic deformation of the individual particles. Examples of this are foams and colloid solutions, e.g., soap foam or a suspension of latex beads in water. The foam bubbles block each other under isotropic (direction-independent) tension, so that the blockage is not canceled out by slight shear stresses. The same applies to colloid solutions when elastic deformation of the contained particles occurs. Foam and colloid solutions can elastically resist a moderate external stress in this way, so they are not ``fragile matter''. Of course, the blockage can also be dissolved in such cases by a stronger stress, which then overcomes the stabilizing  force chain.  

The above examples are characterized by the fact that they involve the interaction of macroscopic structures that are not rearranged by thermal energy but by external forces. In contrast, the dynamics in systems consisting of even smaller particles, namely molecules, will also be temperature-dependent. Here, too, the phenomenon of a continuous transition from liquid to amorphous solid is observed due to a change in density or temperature and is known as the glass transition, which occurs, for example, when a liquid is cooled below its melting point and solidifies amorphously while avoiding crystallization \cite{Cavagna2009a}. Examples of glasses include window glass and various polymers, as well as caramelized sugar or obsidian. Honey or molten rock (lava), on the other hand, are examples of supercooled liquids above the glass transition temperature $T_g$.
 
What thermal and athermal systems have in common is not only the lack of long-range structural order, but also the fact that the dynamics slows down dramatically in the vicinity of jamming until the system finally exhibits properties typical of solids. If, for example, one observes the relaxation of such a system into equilibrium after an external perturbation, the characteristic relaxation times $\tau$ occurring in temperature-driven systems usually do not follow a simple Arrhenius law $\tau(T) = \tau_0\,\exp(\Delta E/(k_BT))$ but seem to diverge at a finite temperature $T_0$:
\begin{equation}
\tau(T) = \tau_0\ \exp\left(\frac{B}{T-T_0}\right)
\label{eq:vft}
\end{equation}
This relationship is known as the Vogel-Fulcher-Tammann equation \cite{Cavagna2009a}.

However, the slowing down of the dynamics in the different systems is subject to different external control parameters: an athermal granular or colloidal system solidifies as a function of increasing packing density or decreasing external mechanical stress. A thermal system, on the other hand, is driven into the glass transition with decreasing temperature or increasing density (i.\,h.\ increasing pressure). A kind of phase diagram can be constructed in which all these phenomena can be uniformly understood as the manifestation of a general jamming transition. This phase diagram, which was initially postulated only speculatively due to the similarity of the phenomena \cite{Liu1998jam}, is increasingly being confirmed experimentally, among other things, because the transition region from thermal to athermal systems has recently also become experimentally accessible \cite{Trappe2001jam}. However, a generally accepted uniform theory of jamming, whether for thermal (glass transition) or athermal systems, or even both together, does not yet exist.

\section{Experimental Method}

\subsection{Photon Correlation Spectroscopy}
\begin{figure}[t!]
\centering
\includegraphics[width=0.7\linewidth]{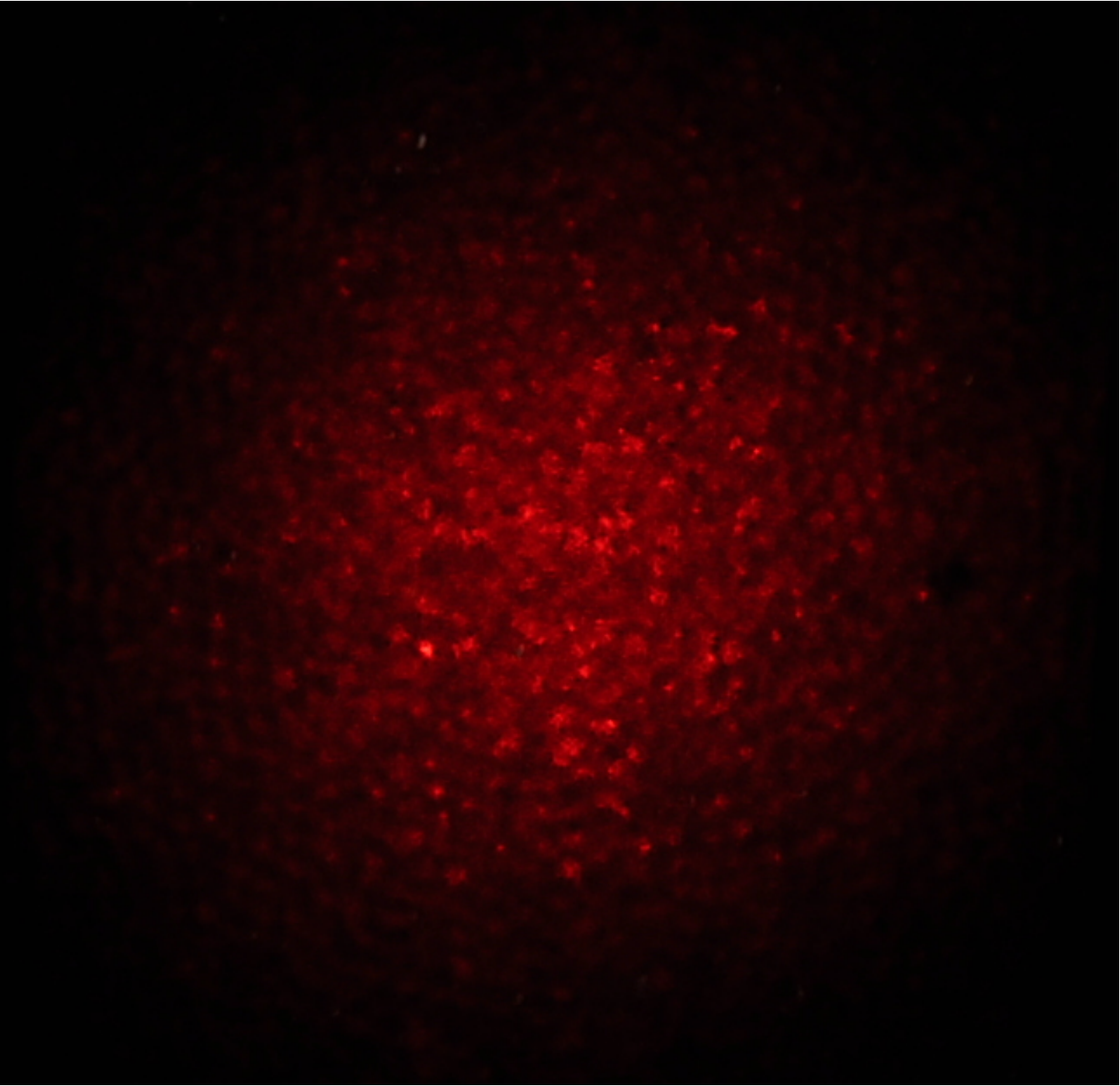}
\caption{Typical speckle pattern of a coherently illuminated disordered system. \cite{M12}}
\label{fig:speckle}
\end{figure}

In order to study the dynamics of granular particles near the jamming transition, photon correlation spectroscopy (PCS) is used in this lab-course experiment. This method is used in many areas of condensed matter physics, for example, to study the dynamics of colloidal particles, polymer solutions and melts, micellar systems, and also small molecules \cite{Pecora1985a}. For grains, which, in contrast to these examples, are not optically transparent, a special variant of the method is used, which is called diffusing wave spectroscopy (DWS)  \cite{Weitz1993a}.

In all versions of the PCS, the time-varying intensity of light scattered by the sample is analyzed. When a coherent light source illuminates a rough surface or a volume with randomly distributed scattering objects, spherical waves of different phases emanate from the scattering centers in the scattering volume, which show a characteristic interference pattern, so-called speckles, in the far field, see Fig.~\ref{fig:speckle}. The spatial intensity distribution due to these phase differences sensitively depends on the arrangement of the scattering centers. Therefore, the time dependent fluctuation of the local light intensity reflects the changing position of the scattering centers in the system under investigation, e.g., the movement of colloidal particles in a suspension. Experimentally, it is crucial that the detected light intensity is not averaged over too large an area, as otherwise the temporal fluctuation is lost. Ideally, therefore, an attempt should be made to restrict the sensitive area of the light detector to the size of approximately a single speckle, which is usually done by using an optical fiber \cite{BP}.

\subsection{Correlation functions in PCS}
In such a PCS experiment the time-varying intensity of light scattered by the sample is analyzed using time correlation functions, a general and important tool in statistical physics to access structural and dynamic properties of many-body systems.  

\begin{figure}[t!]
    \centering
    \includegraphics[width=0.95\linewidth]{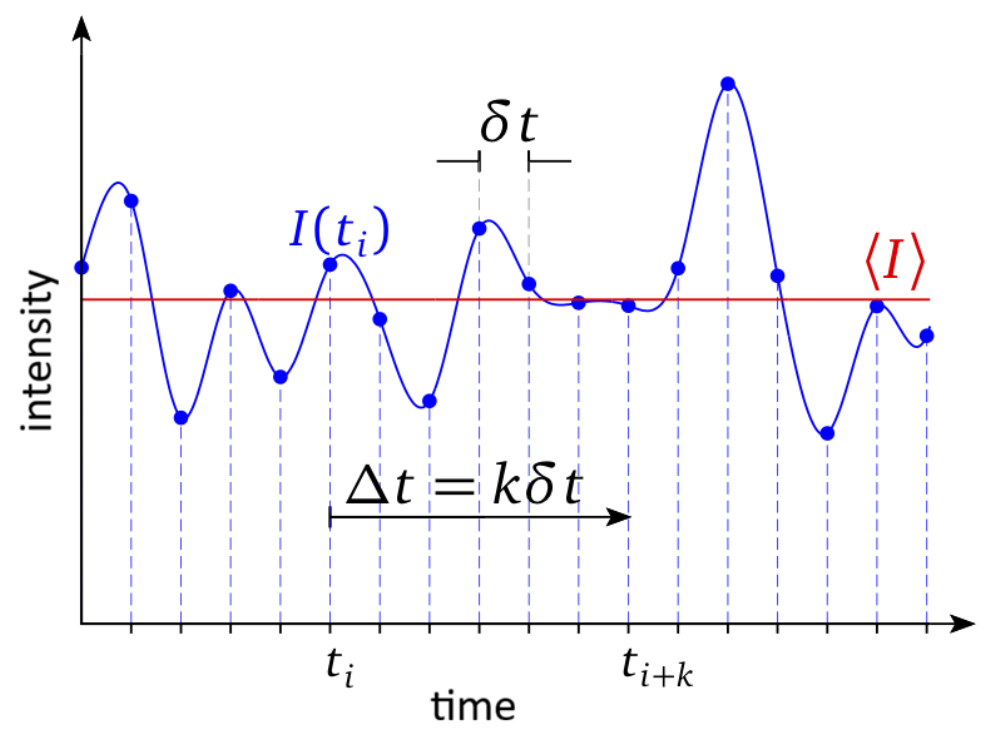}
    \caption{The light intensity fluctuating around the mean value $\langle I\rangle$ as a function of time. The time axis is divided into discrete intervals $\delta t$; an intensity measurement takes place at each time $t_i$. \cite{M12}}
    \label{fig:corr-int}
\end{figure}

In the present context, the time autocorrelation function of the fluctuating scattered light intensity $I(t)$ reads:
\begin{equation}
    g_2(t) = \frac{\langle I(0)\cdot I(t)\rangle}{\langle I\rangle^2}
    \label{eq:g2}
\end{equation}
Here, $\langle\ldots\rangle$ denotes a time average. To be more specific, the fluctuating light intensity is recorded at equidistant time-points $t_i$ as shown in Fig.\ \ref{fig:corr-int}. 
The correlation of the intensities at two times $I(t_0)$ and $I(t_0+\Delta t)$ different by the lag-time $\Delta t = k\cdot\delta t$ is calculated as:
    \begin{equation}
    \begin{split}
    g_2(\Delta t) &  = g_2(k\delta t)  \\&= \frac{\langle I(t_0)\cdot I(t_0+\Delta t)\rangle}{\langle
        I\rangle^2} \\
     & = \frac{1/n\,\sum_i^n I(t_i)\cdot I(t_{i+k})}{\langle I\rangle^2}
    \end{split}
    \label{eq:g2ik}
    \end{equation}
 making use of the fact, that if the system is in a stationary state, the average value of the intensity as well as all correlations are independent of the starting point $t_0$, provided that the number $n$ of values over which the empirical average is taken is sufficiently large. Therefore, it is common practice to set $t_0=0$ in \eqref{eq:g2ik} and to identify the lag-time $\Delta t$ with $t$ so that Eq.\ \ref{eq:g2} results.

The function $g_2(t)$ then drops from the value $\langle I^2 \rangle/\langle I\rangle^2$ at $t=0$ to the value 1, if at long times the intensities $I(0)$ and $I(t)$ have become statistically independent of each other due to the disordered movement of the scattering centers, so that $\langle I(0)\cdot I(t)\rangle=\langle I(0)\rangle\langle I(t)\rangle = \langle I\rangle^2$. Therefore, the quantity $g_2(t) -1$ is often considered instead of $g_2(t)$.

Since the time-dependent position of the scattering centers affects the phase of the scattered wave, the dynamics of the scattering centers is most directly linked to the autocorrelation function of the scattered electric field:
    \begin{equation} 
    g_1(t) = \frac{\langle E^\ast(0)E(t)\rangle}{\langle E^2\rangle} =
    \frac{\langle E^\ast(0)E(t)\rangle}{\langle I\rangle} 
    \end{equation}
The value of this function decreases from
    \begin{equation*}
    \langle E^\ast(0)E(t\to
    0)\rangle/\langle E^2\rangle = 1
    \end{equation*}
for very short times, to a value of
    \begin{equation*}
    \langle E^\ast(0)\rangle\langle E(t)\rangle/\langle E^2\rangle = 0
    \end{equation*}
at very long times. In the simplest case, this relaxation takes place exponentially with a characteristic time $\tau$, i.e:
    \begin{equation}
    g_1(t) = \ee^{-t/\tau}.
    \end{equation}

\begin{figure}[t!]
    \centering
    \includegraphics[width=.5\textwidth]{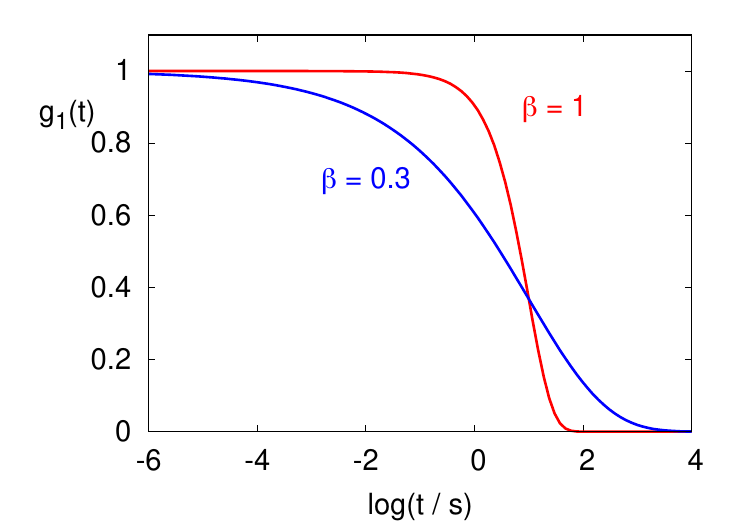}
    \caption{Example of simple and stretched exponential relaxation according to \eqref{eq:kww} with a characteristic time of $\tau = \SI{10}{\second}$ \cite{M12}}
    \label{fig:kww}
\end{figure}
In the vicinity of the jamming transition, however, deviations from this simple behavior are often observed in both thermal and athermal systems, and the correlation decay is described with a Kohlrausch-Williams-Watts function:
    \begin{equation}
    g_1(t) = \ee^{-(t/\tau)^\beta} \qquad \text{mit}\ 0<\beta\leq 1
    \label{eq:kww}
    \end{equation}
For values of $\beta < 1$, such a correlation function is stretched, as illustrated in Fig.\ \ref{fig:kww}. As we shall see below, in our particular experiment, the function $g_1(t)$ is linked to the so-called mean square displacement of the sand grains. In this case, the broadening parameter $\beta$ even provides information on the motional mechanism involved in the dynamics of the particles.  

In order to relate the measured intensity correlation function  to the dynamics of the scattering centers, $g_2(t)$ must be linked to the autocorrelation function of the electric field. This is done using the Siegert relation, which can be derived, if the fluctuations of the electric field are approximately Gaussian distributed.\cite{BP} Then the relation applies:
    \begin{equation}
    g_2(t) = 1 + \Lambda\, |g_1(t)|^2,
    \end{equation}
where $\Lambda$, the so-called coherence factor, which essentially is inversely proportional to the number of speckles over which the intensity is averaged in the experiment, i.e., in an ideal case $\Lambda=1$.

\subsection{Diffusing wave spectroscopy}

Diffusing wave spectroscopy (DWS) is a variant of PCS, in which the multiple scattering of laser light is analyzed, in contrast to conventional PCS experiments in which single scattering is investigated. The method is therefore suitable in principle for turbid, rather ``opaque'' materials such as gels, suspensions, foams, surfactant solutions, or even granulates. The special thing about DWS is that information about the particle dynamics in concentrated samples can be obtained on short length scales, i.e., significantly shorter than the particle diameter, and in a broad time window.

In the theory of DWS, the electric field at the location of the detector is considered, which is composed of contributions from light waves that have traveled different paths $p$ through the sample:
    \begin{equation}
    E(t)=\sum_p E_p \ee^{i\Phi_p(t)}
    \end{equation}
Where $E_p$ is the electric field amplitude and $\Phi_p(t)$ is the associated phase shift caused by the different travel times of light on paths $p$ of different length $s$. The field autocorrelation function thus results in
    \begin{equation}
    \begin{split}
    g_1(t) & = \frac{\langle E(0)\, E^\ast(t)\rangle}{\langle I \rangle} \\&=\sum_p \frac{\langle I_p \rangle}{\langle I \rangle}\langle
        \ee^{-i(\Phi_p(t)-\Phi_p(0))} \rangle \\
     & = \int\langle
        \ee^{-i\,\Delta\Phi_s(t)} \rangle\, P(s)\,\dd s 
    \end{split}
    \label{eq:feld}
    \end{equation}

In the first step, it was assumed that the contributions of individual light paths $p$ to the total intensity are statistically independent of each other. The contribution of a specific path to the total scattering intensity is given by ${\langle I_p \rangle}/{\langle I \rangle}$ and $\sum_p \langle I_p \rangle/\langle I\rangle = 1$. However, since only the path length plays a role for the phase of the electric field, the second step in \eqref{eq:feld} was simplified to an equivalent distribution of path lengths $P(s)$.

It is now also clear why particularly small particle movements can be analyzed with DWS: A change in the total path length by about one wavelength $\lambda$ is caused by a cumulative movement of many particles as scattering centers. It is therefore sufficient for a single particle to move by only a fraction of $\lambda$ to cause a sufficient phase shift in the scattered electric field. Accordingly, DWS can be used to measure the movement of $\mu$m-sized particles on length scales of the order of a few nm.\cite{Weitz1993a}  

The rest of the calculation is now based on an important approximation related to the description of the propagation of light through the scattering medium. Each photon is scattered very often, so that its path through the medium can be represented as a so-called ``random walk'': as a sequence of random steps of length $l^\ast$, which are equally probable in all directions. For a single particle, however, the scattering of light is only isotropic to good approximation if the particle diameter $d$ is much smaller than the light wavelength $\lambda$, i.e. $d/\lambda \ll 1$. Only in this case does the crossing of the mean free path $l$ between two scattering centers lead to a randomization of the scattering direction. Larger particles, on the other hand, usually have a pronounced forward scattering characteristic, so that it takes several scattering events on average until the direction of propagation of the scattered wave is independent of the direction of the incident wave. This length is referred to as the mean transport path length $l^\ast$ and is defined as:
    \begin{equation}
    l^\ast = \frac{l}{1-\langle\cos\theta\rangle},
    \label{eq:transport}
    \end{equation} 
with $\theta$ as the scattering angle and $\langle\dots\rangle$ an average over many scattering events. If the angle $\theta$ is equally distributed, then $\langle\cos\theta\rangle=0$ and $l^\ast=l$. If, on the other hand, the scattering is strongly concentrated in the forward direction, the scattering angle is small on average and $\langle\cos\theta\rangle\approx 1$, then in the limiting case any number of scattering events would be required to randomize the scattering direction and $l^\ast\to\infty$.

\begin{figure}[t]
\centering
\includegraphics[width=0.95\linewidth]{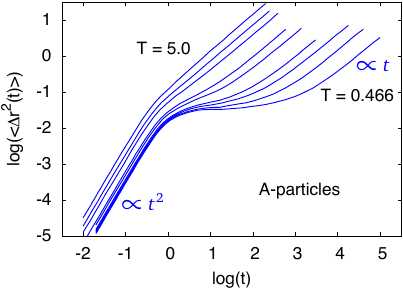}\\
\caption{Example of the time dependence of $\langle\Delta r^2(t)\rangle$, as obtained in the context of molecular dynamics simulations. Here, data from Ref.\ \citenum{Kob1999rev} are shown for a particular type of binary Lennard-Jones system, a model system for supercooled liquids.}
\label{fig:kob}
\end{figure}

As is demonstrated further below, in the simplest case, this results in the field autocorrelation function with $k_0 = 2\pi/\lambda$:
    \begin{equation}
    g_1(t) = \int_0^\infty
    P(s)\,\exp\!\left(-\frac{k_0^2\,s}{3l^\ast}\,\langle\Delta
    r^2(t)\rangle\right)\,\dd s,
    \label{eq:g1msd}
    \end{equation}
where the time-dependent locations of the scattering centers $\vec r_i(t)$ enter as the mean square displacement (MSD): $\langle\Delta r^2(t)\rangle = 1/N\,\sum_i\langle(\vec r_i(t) - \vec r_i(0))^2\rangle$. This is the central quantity analyzed in the DWS and a parameter that is often calculated in molecular dynamics simulations, for example. It not only provides information about how far a particle moves on average in a certain time interval, but also in what way: If the particle does not experience any collisions with other particles on its trajectory, i.e., it is a ballistic, ``free flight'' kind of motion, then $\langle\Delta r^2(t)\rangle = \langle \delta v^2\rangle t^2$. Here, $\langle \delta v^2\rangle$ is the mean square fluctuation of the velocity and is therefore a measure of the kinetic energy contained in the system. It can be used to define a so-called granular temperature. If, on the other hand, a particle experiences many collisions and therefore,  on average, performs a random walk, this can be represented as a diffusive process with a diffusion constant $D$ after a sufficient number of steps: $\langle\Delta r^2(t)\rangle = 6D\,t$. In both cases, $\langle\Delta r^2\rangle$ grows as a function of time without limitation. If, on the other hand, the dynamics of a particle is subject to a geometric restriction, either imposed from the outside (e.\,B.\ moving particle in a porous, solid matrix) or by the collective effect of many other particles (``cage effect'' in supercooled liquids), then $\langle\Delta r^2(t)\rangle$ increases in a sub-diffusive manner and approaches a constant value at long times and thus indicates a localization of the particles on average. A typical example of  $\langle\Delta r^2(t)\rangle$, as observed in simulations and experiments not only in molecular but also in colloidal and granular systems, is shown in Fig.~\ref{fig:kob}.

To obtain the MSD from the field autocorrelation function, strictly speaking the distribution of path lengths $P(s)$ traveled by the light through the sample must be known, and \eqref{eq:g1msd} shows that long photon paths contribute to a faster drop in the correlation function. In order to calculate $P(s)$, the propagation of light is described in the diffusion approximation, i.\,e.\ a diffusion equation for the energy density of the light wave must be solved. One uses the fact that \eqref{eq:g1msd} has the form of a Laplace transform and the solution turns out to depend on the exact geometry of the sample and the optics used \cite{Pine1988,Weitz1993a}. Usually, the resulting expression must then be inverted numerically to obtain the MSD. However, it can be seen that the correlation function is dominated by an average path length $s_0$, especially in transmission geometry: If we assume that the path of the light through the sample can be described as a random walk with an average of $N_R$ steps of length $l^\ast$, then the mean path length is given by $s_0 = N_R\cdot l^\ast$ and the mean end-to-end distance of the random walk is $L^2 = N_R\cdot l^{\ast 2}$. If one now assumes that in transmission geometry the cell diameter just corresponds to the mean end-to-end distance of the random walk, the estimate for the mean path length traveled by the light is $s_0 = L^2 / {l^\ast}$. Therefore, the simplest assumption for $P(s)$ is that of a $\delta$ distribution around $s_0$, i.e:
    \begin{equation}
    P(s)\simeq \delta(s-s_0)
    \label{eq:ps0}
    \end{equation}
and \eqref{eq:g1msd} becomes:
    \begin{equation}
    g_1(t) = \exp\left({-\frac{k_0^2}{3}\,  \left(\frac{L}{l^\ast}\right)^2\, \left\langle \Delta r^2(t) \right\rangle}\right) 
    \label{eq:g1msd_simp}
    \end{equation}
Of course, as said, determining $P(s)$ by solving the diffusion equation for scattered light would be the more accurate way, leading to much more involved expressions.\cite{Pine1988, Weitz1993a} Thus, for the purpose of the present lab course experiment we prefer to stay with the simplest approximation \eqref{eq:g1msd_simp}. 

\subsection{The DWS correlation function for a vertically agitated granular medium}
To start, the equivalent of Eqs.\ \eqref{eq:g1msd} and \eqref{eq:g1msd_simp} for a vertically shaken granular bed should be deduced. As already stated, the signal at the detector location is generated by a superposition of light waves that have traveled different paths $p$ and starting from \eqref{eq:feld} we obtain:
    \begin{equation}
    g_1(t) = \sum_p \frac{\langle E_p(0)\, E_p^\ast(t)\rangle}{\langle I \rangle} =\sum_p \frac{\langle I_p \rangle}{\langle I \rangle}\langle \ee^{-i\,\Delta\Phi_p(t)} \rangle 
    \label{eq:feld2}
    \end{equation}

\begin{figure}[t!]
\centering
\includegraphics[width=0.95\linewidth]{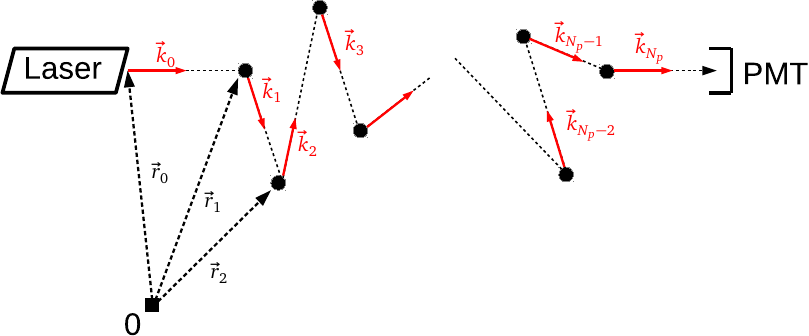}\\
\caption{
Schematic representation of a possible light path $p$ from the laser ($i=0$: $\vec{r}_0$) to the detector ($i=N_p+1$: $\vec{r}_{N_p+1}$) through the sample, with the position vectors of the individual scattering centers $\vec r_i$, $i=1\dots N_p$ and the wave vectors $\vec k_i$ emanating from the scattering centers. \cite{M12}
}
\label{fig:lichtweg}
\end{figure}

The aim is now to analyze the phase shift $\Delta\Phi_p(t)$ arising on the path $p$ more in detail. One possible light path through the sample is shown schematically in Fig.~\ref{fig:lichtweg}. As each scattering event is elastic in good approximation, all wave vectors have the same magnitude $k_i = 2\pi/\lambda$, but different directions.
    \begin{equation}
    \Phi_p(t)=\sum_{i=0}^{N_p} \vec{k}_i(t)[\vec{r}_{i+1}(t)-\vec{r}_{i}(t)]
    \label{eq:phi}
    \end{equation}
The positions of the laser and detector $\vec{r}_{0}$ and $\vec{r}_{N_p+1}$ are fixed, all other $i=1,\dots,N_p$ are time-dependent. To introduce vertical shaking motion, the position $\vec{r}$ at time $t$ is made up of the periodically changing position of the cuvette in the $z$ direction and the relative position of a grain of sand within the cuvette $\vec x_i(t)$: 
    \begin{equation}
    \vec{r}_i(t)=A_0 \hat{e}_z \sin(\omega(t+t_0))+\vec{x}_i(t)
    \label{eq:comp}
    \end{equation}
Here, $t_0$ refers to the initial position of the cuvette at time $t=0$. \eqref{eq:phi} gives the phase shift on the path $p$:
    \begin{eqnarray}
    \Delta\Phi_p(t) & = &\Phi_p(t)-\Phi_p(0) \nonumber \\
    & = & \sum_{i=0}^{N_p}
    \vec{k}_i(t)[\vec{r}_{i+1}(t)-\vec{r}_{i}(t)] \nonumber\\ & &-\sum_{i=0}^{N_p}
    \vec{k}_i(0)[\vec{r}_{i+1}(0)-\vec{r}_{i}(0)] \nonumber\\ 
    &= & \sum_{i=0}^{N_p}
    \vec{k}_i(0)[\Delta\vec{r}_{i+1}(t)-\Delta\vec{r}_{i}(t)] \nonumber\\ & &+\sum_{i=0}^{N_p}
    \Delta\vec{k}_i(t)[\vec{r}_{i+1}(t)-\vec{r}_{i}(t)], \label{eq:dphi1}
    \end{eqnarray}
with $\Delta\vec r_i(t) = \vec r_i(t) - \vec r_i(0)$ and $\Delta k_i(t)= \vec k_i(t) - \vec k_i(0)$.

Now we assume that in first approximation $\Delta\vec{k}_i(t)\perp(\vec{r}_{i+1}(t)-\vec{r}_{i}(t))$ (see Fig.~\ref{fig:lichtweg}, keeping in mind the constant magnitude of the wave vector and also that $\Delta r_i(t)\ll |\vec{r}_{i+1}(t)-\vec{r}_{i}(t)|$ is a small shift!) and so the second term in \eqref{eq:dphi1} is negligible. So, what remains is:
    \begin{equation}
    \Delta\Phi_p(t)=\sum_{i=0}^{N_p} \vec{k}_i(0)[\Delta\vec{r}_{i+1}(t)-\Delta\vec{r}_{i}(t)]
    \label{eq:deltaphi}
    \end{equation}
Now we insert \eqref{eq:comp} into \eqref{eq:deltaphi} and realize that all oscillatory components within the cuvette cancel and only the components at the edges remain, i.e:
    \begin{eqnarray}
    \Delta\Phi_p(t) & = & \vec{k}_0A_0 \hat{e}_z[
    \sin(\omega(t+t_0))-\sin(\omega t_0)]\nonumber \\&&-\vec{k}_{N_p}A_0 \hat{e}_z[
    \sin(\omega(t+t_0))-\sin(\omega
    t_0)] \nonumber \\ 
    & &
    +\sum_{i=0}^{N_p-1}\vec{k}_i\,\Delta\vec{x}_{i+1}(t)-\sum_{i=1}^{N_p}\vec{k}_i\,\Delta\vec{x}_{i}(t)\nonumber
    \\
    & = & (\vec{k}_0-\vec{k}_{N_p})A_0 \hat{e}_z[
    \sin(\omega(t+t_0))-\sin(\omega
    t_0)]\nonumber \\ & & +\sum_{i=1}^{N_p}\vec{k}_{i-1}\,\Delta\vec{x}_{i}(t)-\sum_{i=1}^{N_p}\vec{k}_i\,\Delta\vec{x}_{i}(t)
    \end{eqnarray} 
With the definitions $\vec{q}_i=\vec{k}_i-\vec{k}_{i-1}$ and $\kappa_p = (\vec{k}_{N_p}-\vec{k}_0)\cdot\hat{e}_z = \vec{k}_{N_p}\cdot\hat{e}_z$ (because $\vec{k}_0\perp\hat{e}_z$, while $\vec{k}_{N_p}\cdot\hat{e}_z$ depends on the path $p$) we obtain:
    \begin{equation}
    \begin{split}
    \Delta\Phi_p(t) = & -A_0\kappa_p\,[ \sin(\omega(t+t_0))-\sin(\omega t_0)] \\ &- \sum_{i=1}^{N_p}\vec{q}_i\Delta\vec{x}_i(t)
    \end{split}
    \label{eq:delphi}
    \end{equation}
Now, according to the central limit theorem, the phase shifts $\Delta\Phi_p$ along different paths are approximately Gaussian distributed. Therefore, $\langle\Delta\Phi_p(t)\rangle = 0$ applies and the odd moments of $\Delta\Phi_p$ do not contribute in \eqref{eq:feld2}. Thus, the average for the phase factor in \eqref{eq:feld2} can be written as:
    \begin{equation}
    \left\langle e^{-i\Delta\Phi_p(t)}\right\rangle = 
    \sum_{m=0}^\infty \left\langle\frac{(-i\Delta\Phi_p(t))^m}{m!} \right\rangle = \ee^{-\langle\Delta\Phi_p^2(t)\rangle/2}
    \label{eq:phigauss}
    \end{equation}
In the last step we used that for Gaussian random variables all higher central moments can be calculated from the second moment according to: $\langle\Delta\Phi^{2m}\rangle = (2m)!/(2^m m!) \langle\Delta \Phi^2\rangle^m$.

Now we note that in \eqref{eq:delphi} the oscillatory component $A_0 \sin\dots$ and the displacement of the sand grains $\Delta\vec{x}_i(t)$ are statistically independent, as are the individual $\vec{q}_i \Delta\vec{x}_i$ and the respective $q_i$ and $x_i$. In addition, we have $\langle\Delta x_i\rangle=0$ and $\langle\kappa_p\rangle=0$. Therefore, we obtain:
    \begin{eqnarray}
    \langle\Delta\Phi_p^2(t)\rangle & = & \langle\kappa_p^2\rangle
    A_0^2[\sin(\omega(t+t_0))-\sin(\omega
    t_0)]^2\nonumber \\&&+\sum_i\langle(\vec{q}_i\Delta\vec{x}_i(t))^2\rangle
    \nonumber \\ 
    & = & \kappa^2 A_0^2 [\sin(\omega(t+t_0))-\sin(\omega
    t_0)]^2 + \nonumber\\
    && + N_p\langle q_x^2\rangle\langle\Delta x_x^2(t)\rangle +
    N_p\langle q_y^2\rangle\langle\Delta x_y^2(t)\rangle \nonumber \\&& +  N_p\langle q_z^2\rangle\langle\Delta x_z^2(t) \rangle\nonumber\\
    & = & \kappa^2 A_0^2 [\sin(\omega(t+t_0))-\sin(\omega
    t_0)]^2 \nonumber \\&&+ 3 N_p\, \frac{\langle q^2 \rangle}{3}\, \frac{\langle \Delta
        x^2(t)\rangle}{3}
    \end{eqnarray}
Now, the mean square scattering vector for a given scattering event is given by
    \begin{equation}
    \langle
    q^2 \rangle= \langle(2 k_0 \sin(\theta/2))^2\rangle= 2 k_0^2
    \langle 1-\cos{\theta}\rangle= 2 k_0^2 \frac{l}{l^\ast}
    \end{equation}
and is thus determined according to the definition of \eqref{eq:transport} by the ratio of the mean free scattering path length $l$ to the mean free transport path length $l^\ast$. Using the total length of the path $s_p = N_p\cdot l$ therefore results in
    \begin{equation}
    \begin{split}
    \langle\Delta\Phi_p^2(t)\rangle  = & \kappa^2 A_0^2
    [\sin(\omega(t+t_0))-\sin(\omega t_0)]^2 \\ & +\frac{2}{3}
    k_0^2 \frac{s_p}{l^\ast} \langle \Delta x^2(t)
    \rangle 
    \end{split}
    \end{equation}

Thus, the phase shift depends only on the path length and not on other properties of the light path and so in \eqref{eq:feld2} we can move from the contribution of a single path to the total scattering intensity $\langle I_p\rangle/\langle I\rangle$ to a distribution of path lengths $P(s)$, as already anticipated in \eqref{eq:feld}:
    \begin{equation}
    g_1(t)= \int_0^{\infty} P(s) \exp[-\frac{1}{2}\langle \Delta
    \Phi_s^2(t) \rangle] \dd s 
    \end{equation}
Now we choose the simplification $P(s)\simeq \delta(s-s_0)$ according to \eqref{eq:ps0} with $s_0=L^2/l^\ast$ and obtain the simplest case for the field autocorrelation function under vertical excitation as:
    \begin{equation}
    \begin{split}
    g_1(t,t_o)  = & \ee^{-\frac{1}{2} \kappa^2 A_0^2
        [\sin(\omega(t+t_0))-\sin(\omega t_0)]^2}\ \\ &  \ee^{-{k_0^2}/{3}\,
        ({L}/{l^\ast})^2\, \langle \Delta x^2(t) \rangle}
    \end{split}
    \label{eq:g1t0}
    \end{equation}

\begin{figure}[t]
\includegraphics[width=0.95\linewidth]{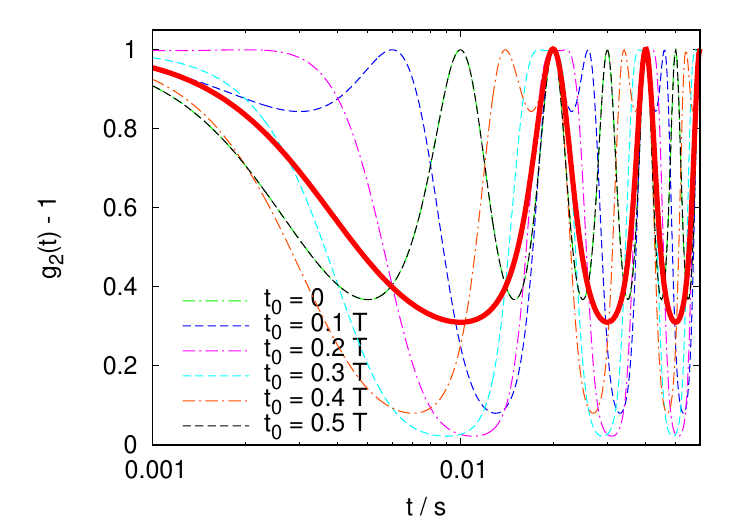}
\caption{The oscillatory part of the intensity correlation function $g_2(t)-1$ according to \eqref{eq:g2int} (solid line) for $\omega/2\pi = \SI{50}{\hertz}$, as well as individual contributions to the integral resulting from different start times $t_0$ (dashed).}
\label{fig:dws-sin}
\end{figure}

This expression still contains the initial value of the oscillatory part of the phase shift. As the intensity of the light is averaged over time in the real experiment, \eqref{eq:g1t0} must be averaged over all initial states of the cuvette oscillation $t_0$, i.e:
    \begin{equation} 
    g_2(t)-1=1/T\int_0^T |g_1(t,t_0)|^2\,\dd t_0
    \end{equation}
and so the result is:
    \begin{equation} 
    \begin{split}
    g_2(t)-1 & = \exp(-2/3\, k_0^2(L/l^\ast)^2\langle\Delta
    x^2(t)\rangle)  \cdotp 1/T \\
     &   \int_0^T \exp(-\kappa^2A_0^2[\sin(\omega(t+t_0))-\sin(\omega
    t_0)]^2)\, \dd t_0
    \end{split}
    \label{eq:g2int}
    \end{equation}
In Fig.~\ref{fig:dws-sin}, the integral of \eqref{eq:g2int} was calculated numerically and is shown together with some contributions resulting from different initial times $t_0$. Obviously, the vertical vibration shows in the correlation function as periodic "echoes" of constant height, as long as the prefactor in front of the integral, which contains the mean square displacement of the grains, does not reduce the echo amplitude.
\begin{figure}[t]
    \includegraphics[width=0.95\linewidth]{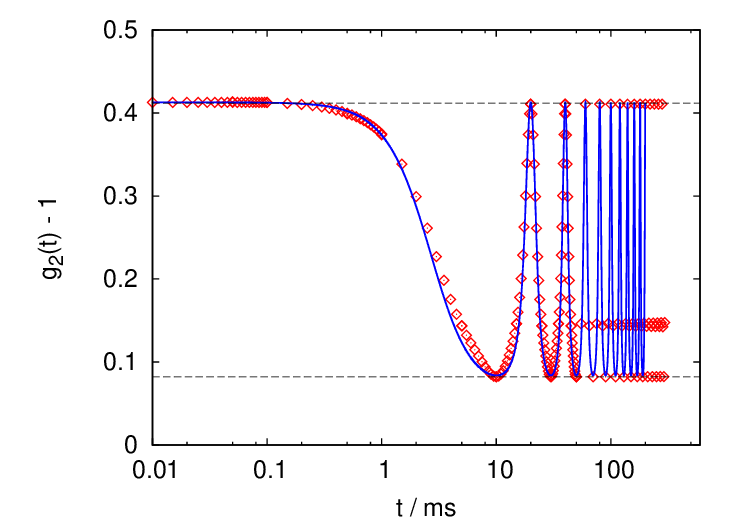}
    \caption{Styrofoam measurement used as a static multiple-scattering sample to determine the oscillatory correlation function.}
    \label{fig:1}
\end{figure}

\section{Experimental Details}
Fig.\ \ref{fig:aufbau} schematically shows the setup of the experiment. A HeNe laser at $\lambda=\SI{632.8}{\nano\meter}$ irradiates a cuvette ($L=\SI{1}{\centi\meter}$, $B=\SI{5}{\centi\meter}$, $H=\SI{12}{\centi\meter}$) which is filled to a height of approximately $\SI{8}{\centi\meter}$) with small glass beads ($d=(315\pm 15)\,\si{\micro\meter}$). The cuvette is located on a vibration unit that works according to the loudspeaker principle (elastic membrane, magnet, coil) and is controlled by a function generator. The guide rods at the upper end of the cuvette ensure the stability of the system against lateral movements so that the cuvette vibrates stably in the $\hat e_z$ direction with an amplitude $A_0$ and a circular frequency $\omega$: $\vec r(t) = A_0\,\hat e_z\sin(\omega t)$. A capacitive acceleration sensor is attached directly to the surface of the cuvette. This sensor converts the instantaneous acceleration experienced by the system into a voltage that can be displayed directly on an oscilloscope in units of gravitational acceleration $g$ in two measurement ranges of $600\,\si{\milli\volt}/g$ or $300\,\si{\milli\volt}/g$, so that the dimensionless acceleration amplitude $\Gamma = \omega^2\,A_0/g$ can be directly obtained by reading the sensor voltage.

The scattered light is coupled into an optical single-mode fiber. This type of optical fiber is characterized by the fact that its core diameter is so small (only a few wavelengths) that the light can propagate only parallel to the central fiber axis and not by total reflection on the fiber wall. In our experiment, we used this property of the fiber to select a single speckle from the scattered light because light is only collected through the fiber from a small range of solid angles. The fiber then couples the light into a photomultiplier tube (PMT), which sends one electrical pulse per detected photon to a fast counter/timer card.

\section{Results and Discussion}
\begin{figure*}[tbh]
\includegraphics[width=0.49\linewidth]{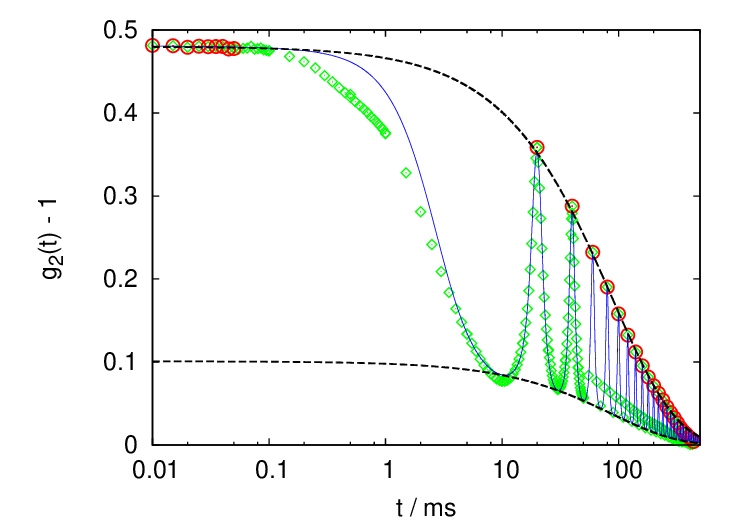}
\includegraphics[width=0.49\linewidth]{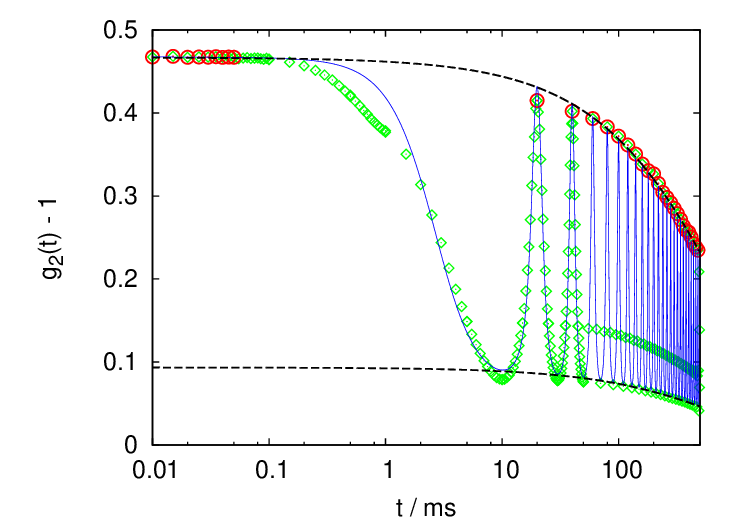}
\caption{Intensity correlation functions of \SI{315}{\micro\meter} glass spheres vibrated at a frequency of \SI{50}{\hertz} and acceleration amplitudes of $\Gamma=0.95$ (left) and $\Gamma=0.88$ (right), respectively.}
\label{fig:2}
\end{figure*}
To first check the result of the oscillatory part calculated in \eqref{eq:g2int} and Fig.\ \ref{fig:dws-sin} a styrofoam block provides strong multiple scattering and is therefore well-suited as a static reference sample. The resulting intensity correlation function is shown in Fig.~\ref{fig:1}. The function given by Eq.~\ref{eq:g2int} is shown in blue for an oscillation frequency of \SI{50}{\hertz} and assuming a static mean squared displacement (MSD). The parameter $\kappa^2A_0^2=1.45$ was obtained by a least squares fit procedure and the resulting function reproduces the oscillatory behavior of the correlation function very well, as expected for a static system where no structural rearrangements occur. We mention that on a logarithmic time axis, the spacing between successive echoes diminishes at long times.

When the glass beads are subjected to vertical vibration, particle rearrangements occur, resulting in the decay of the correlation function. Two representative intensity correlation functions $g_2$ for \SI{315}{\micro\meter} glass spheres at $\Gamma=0.95$ and $\Gamma=0.88$ and an excitation frequency of \SI{50}{\hertz} are shown in Fig.~\ref{fig:2}. In contrast to the static case, the correlation function now exhibits a decay envelope due to particle motion. Assuming that particle dynamics follows the stretched exponential given by Eq.~\ref{eq:kww}, the measured correlation functions can be well described when curve-fitting the oscillation maxima (or minima) with a stretching parameter of $\beta=0.8$. What can also be seen is that the higher the energy input into the system, i.e., the larger the value of $\Gamma$, the faster the observed relaxation.

\begin{figure}[t]
\includegraphics[width=0.95\linewidth]{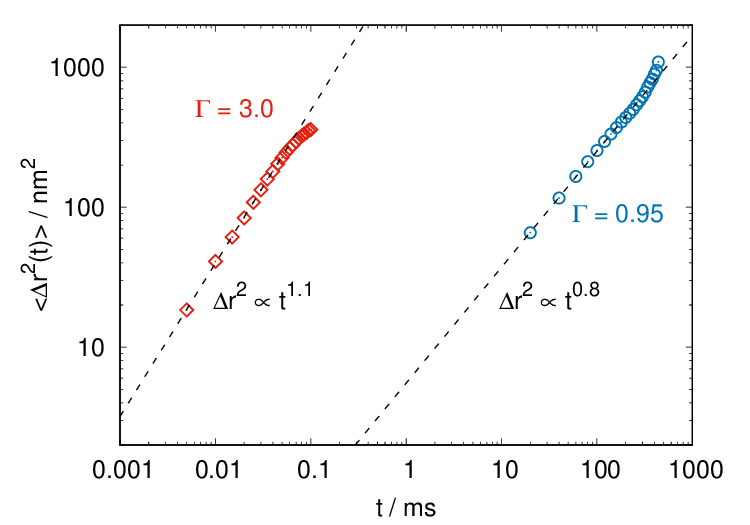}
\caption{Mean squared displacements calculated from the obtained correlation functions.}
\label{fig:3}
\end{figure}

Using \eqref{eq:g2int} the mean squared displacements can be determined either from the maxima of the correlation function at long times and small acceleration amplitudes or from the initial decay at large amplitudes, where significant displacements occur on timescales faster than the vibration cycle. In both cases, the transport mean free path $l^\ast$ has to be known to apply \eqref{eq:g2int} and obtain the MSD. Actually, $l^\ast$ can be accessed using transmission measurements and employing well-chosen reference samples.\cite{Leutz1996a} For our present purpose, we simply estimate $l^\ast$ based on experimental literature data for similar systems\cite{Zivkovic2011a} as $l^* \approx 2\,d$, where $d$ is the particle diameter. The resulting MSDs are shown in Fig.~\ref{fig:3} for $\Gamma = 3$ and $\Gamma = 0.95$. The extracted power-law exponents are close to unity in both cases, therefore approximately diffusive particle motion is observed. The deviations suggest a tendency towards hyperdiffusive behavior at large acceleration and slightly subdiffusive motion at lower $\Gamma$. The corresponding mean square displacement amplitudes indicate motions on the order of some \SI{10}{\nano\meter}, which is small compared to the particle diameter.

\begin{figure*}[t]
\includegraphics[width=0.49\linewidth]{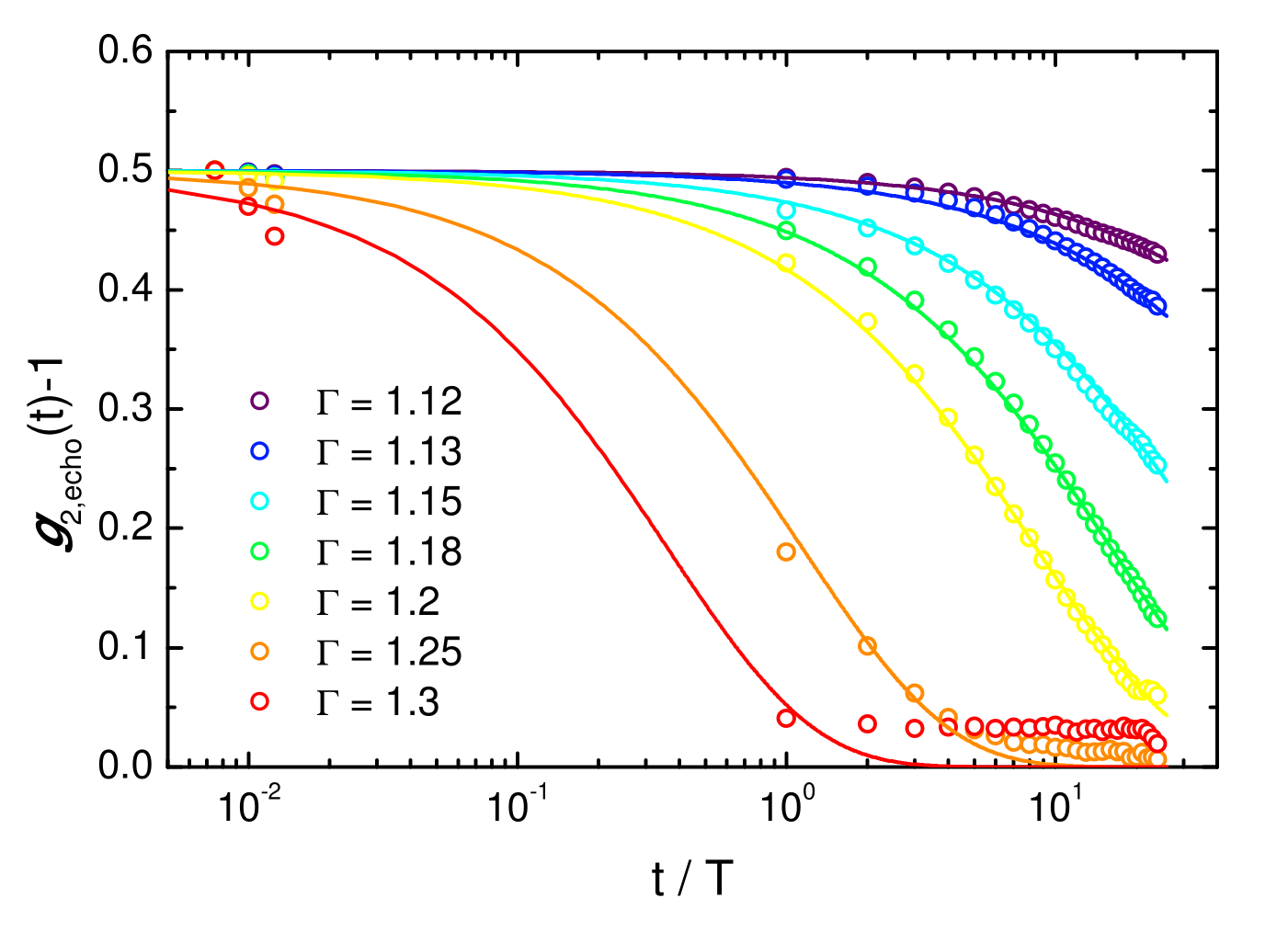}
\includegraphics[width=0.49\linewidth]{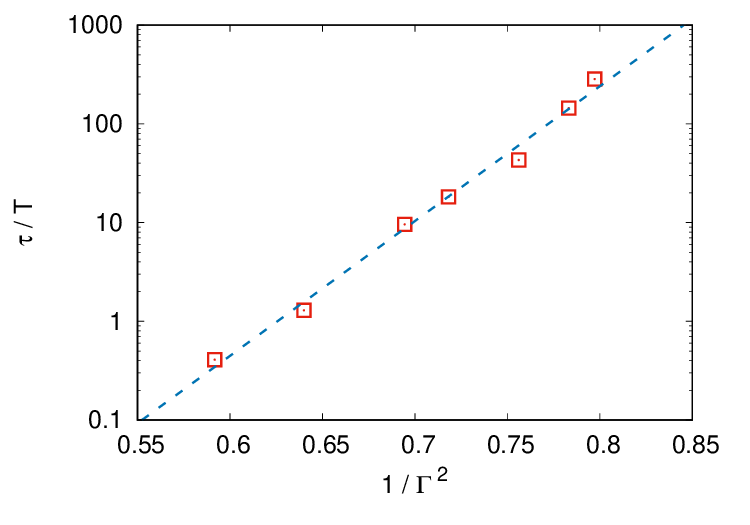}
\caption{Left: Echo maxima of Normalized relaxation times $\tau/T$ at an excitation frequency of \SI{50}{\hertz} as a function of inverse squared dimensionless acceleration for different excitation amplitudes. The data approximately follow "Arrhenius"-like behavior.}
\label{fig:4}
\end{figure*}
As already discussed in the introduction, it is common practice for thermally driven systems to consider relaxation behavior in terms of an Arrhenius law, or in more complex cases, deviations therefrom with a Vogel-Fulcher equation, see \eqref{eq:vft}. In athermal systems like granular materials, the situation is fundamentally different due to the dissipative nature of interparticle collisions, but especially in dense systems close to the jamming transition, similar concepts are often applied.\cite{Liu1998jam} One example is the correlation of the dynamic slowing down close to jamming with a decrease in granular temperature. The latter is defined considering the fact that continuous energy input results in a random motion of particles, which is quantified by the mean-square value of velocity fluctuations $\langle\delta v^2\rangle$, to which the granular temperature is proportional.\cite{Biggs2008a} 

In case of a vertically vibrated granular bed, it was demonstrated in a previous study by Zivkovic et al.\cite{Zivkovic2011a} that on sinusoidal excitation the granular temperature is approximately proportional to the square of the peak vibrational velocity, which is given by $v_p^2 = \omega^2 A_0^2$. Thus, in order to arrive at an "Arrhenius-style" plot for our granular medium, the excitation amplitude $A_0$ is varied and the average relaxation times $\tau$ are extracted from a curve-fit of the echo amplitudes with a KWW stretched exponential function, see Fig.\ \ref{fig:4}(left). Again, all measurements were performed at the same \SI{50}{\hertz} excitation frequency, so that when changing the amplitude $A_0$ the squared dimensionless acceleration becomes a measure of the granular temperature: $\Gamma^2 \propto v_p^2=\omega^2A_0^2$. Thus, in Fig.~\ref{fig:4} the extracted time constants are normalized by the oscillation period $T$ and shown as a function of $1/\Gamma^2$ at a fixed excitation frequency. The observed straight line indicates that the relaxation times indeed follow an Arrhenius-like dependence,  suggesting that close to the jamming transition particle rearrangements are governed by activated jumps over energy barriers similar to the behavior of thermal systems close to the glass transition. Whether an even stronger, VFT-like dependence on granular temperature would be recovered if, e.g., a larger range of granular temperatures or energy inputs was considered remains an open question.

\section{Conclusion}
In conclusion, we have presented a comparatively simple setup that is used at the undergraduate physics lab course at TU Darmstadt to investigate jamming transition in vertically vibrated granular systems. It turns out that vibrations provide an effective and tunable means of injecting controlled amounts of energy into a granular bed. Describing this energy in terms of $1/\Gamma^2$ provides an analogy to thermodynamic descriptions known from supercooled liquids. Within this framework, also current research shows that scattering theory proves to be well-suited for systems of glass beads, similar to the current approach; although additional scattering mechanisms may contribute, the approach can be used to access mean square displacements of particles like it was demonstrated in granular polystyrene \cite{heitmeier2026vibrated,kunzner2025dynamics}.

However, significant challenges arise upon further densification at decreasing granular temperature. Compaction becomes increasingly slow below $\Gamma$ one compared to the observation times accessible in this work, leading to pronounced hysteresis effects. As a consequence, the system loses ergodicity and cannot be considered in a steady state. Under these conditions, alternative techniques such as camera-based photon correlation spectroscopy (PCS) \cite{gabriel2018depolarized,mayo2025observing} must be considered to properly account for non-ergodic dynamics. Related experiments by Kunzner et al.\ \cite{kunzner2025dynamics} have already demonstrated approaches to track the time-dependent evolution of compaction.

Moreover, densification in fluidized beds is inherently heterogeneous. Due to the particle weight, pressure distributions resemble those in silos and are only approximately homogeneous in the central regions\cite{sperl2006experiments,janssen1895versuche}. Environmental factors further complicate the situation: humidity strongly influences inter-particle cohesion through clustering and charge redistribution.\cite{Schella2017a} Vibrated granular systems tend to accumulate electrostatic charge, which can significantly alter their behavior, while humidity is required to enable charge dissipation.\cite{Zhao2023a}

Achieving truly dense states, therefore, requires access to very low effective granular temperatures, where the gravitational energy barrier has to be lowered below the kinetic energy. On Earth, this can be realized through drop tower experiments  or parabolic flights.\cite{born2017dense} For slow dynamical processes, where the duration of the reduced/zero gravity condition  is crucial, experiments in microgravity environments such as the ISS are particularly advantageous, although experimentally demanding. Under such conditions, it becomes possible to maintain dense yet fluid-like states at low granular temperatures over extended periods of time.\cite{mayo2025observing,Born2021mug} While the presented experiment does not tackle the problem of gravity in granular systems, the setup is easy to build and simple in operation, and thus provides access to this fascinating field of current research on the level of a lab course experiment.

\bibliography{m12-paper}

\clearpage
\end{document}